\documentclass{ws-p8-50x6-00}
\usepackage{psfig}
\begin{document}

\title{ Astrophysical Fractals: Interstellar Medium and Galaxies }
\author{F. Combes}

\address{DEMIRM, Observatoire de Paris, 61 Av. de l'Observatoire,
F-75 014, Paris, FRANCE\\E-mail: bottaro@obspm.fr}


\maketitle

\abstracts{
 The interstellar medium is structured as a hierachy
of gas clouds, that looks self-similar over 6
orders of magnitude in scales and 9 in masses. This is one of
the more extended fractal in the Universe.
At even larger scales, the ensemble of galaxies
looks also self-similar over a certain ranges of
scales, but more limited, may be over
3-4 orders of magnitude in scales. These two fractals
appear to be characterized by similar
Hausdorff dimensions, between 1.6 and 2.
The various interpretations of these structures
are discussed, in particular formation theories based
on turbulence and self-gravity. In the latter,
the fractal ensembles are considered in a critical
state, as in second order phase transitions,
when large density fluctuations are observed, that 
also obey scaling laws, and look self-similar over
an extended range.}

\bigskip
\section{Introduction}

 Fractals are ensembles that can 
be defined by their self-similarity. The name
has been introduced by Mandelbrot (1975) to define 
geometrical or mathematical sets, that have a non-integer,
i.e. fractional dimension. The dimension determines
whether a system is homogeneous, and what fraction
of space is filled. For a homogeneous density, 
the mass of the medium is increasing as the 3rd power
of its radius (in 3D), while when the medium is
fractal, he may occupy a tiny fraction of space, 
and the mass contained within a scale $r$ is M$\propto r^D$,
with $D$, the Hausdorff dimension, lower than 3.

The two fractals described here, the interstellar
medium (ISM) and galaxies, have a dimension
$D \approx 1.7$. Of course these physical ensembles
are only approximations of mathematical fractals.
They are self-similar only between two limiting scales,
where boundary effects occur, while a pure
mathematical fractal is infinite; and they
are quite randomly distributed, their
self-similarity being only statistical.

\bigskip
 \section{The Interstellar Medium}

The gaseous interstellar medium has a very irregular and fragmented
structure. It consists of clouds of hydrogen, either atomic
or molecular, according to its density or column density. 
The atomic gas is less dense and more diffuse in general,
while the molecular gas gathers the most clumpy, cold
and dense phase (see the survey of the Milky Way in Molecular Clouds
by Dame et al 1996).

\begin{figure}[t]
\psfig{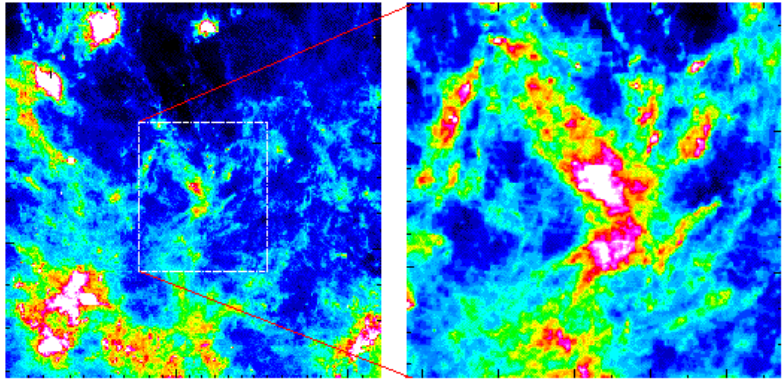}
\caption{{\it Left:} IRAS 100 $\mu$m map of the Taurus molecular cloud complex,
traced by the dust emission. The square is $\sim$ 4000 pc$^2$.
{\it Right} Zoom of the central region (the 
square is now  $\sim$ 400 pc$^2$).}
\label{taurus}
\end{figure}

At any scale, the self-similar appearance of the clouds
ensemble makes it difficult to
determine their absolute scale on photographs, without any
other information (about velocity, distance, etc..).
In Fig \ref{taurus}, we show for example maps of the nearby
Taurus cloud, observed in the far-infrared at 100$\mu$m
with the IRAS satellite. The emission comes from dust
heated by nearby stars, or by the interstellar radiation field.
 The right-hand side map is an enlarged view of the central left map,
but still quite similar. Constraints due to spatial resolution
of telescopes, cannot in general allow to
observe more than a few orders of magnitude in scale range
for the same cloud, but the observation of several clouds
at various distances in the Milky Way (from 0.05 to 15 kpc) or
in external galaxies, have established the scaling laws over
a wider range.

\bigskip
\subsection{ Low and High-mass Cut-off for the ISM Fractal}

The largest self-gravitating entities in the Galaxy are the
 so-called Giant Molecular Clouds (or GMC) of about 100pc
diameter, and 10$^6$ M$_\odot$ in mass. Larger clouds
cannot exist since they would be teared off by the galactic shear,
i.e. the tidal forces due to the galactic potential itself.
This is the high cut-off scale in the fractal structure.
What is the smallest size?

It is difficult to observe directly in emission the smallest structure,
due to lack of spatial resolution and sensitivity. But structures
of about 10-20 AU in size (i.e. $\sim$ 10$^{-4}$ pc) have
been observed for a long time through scattering of the quasar
light (Fiedler et al. 1987, 1994, Fey et al. 1996): clumps in the electronic
density diffract the light rays from remote quasars, and produce
an "extreme scattering event" (ESE) lasting for a few months,
in their rapid motion (100-200km/s) just in front of the quasar.
QSOs monitoring during several years has determined that the number
of scattering structures is 10$^3$ times as numerous as stars 
in the Galaxy. The problem of stability and life-time
of these structures, with much higher pressure than surroundings,
can be solved if they are self-gravitating (Walker \& Wardle 1998);
they are then of 10$^{-3}$ M$_\odot$ in mass, and have a gas density
around  10$^{10}$ cm$^{-3}$. They correspond to the smallest
fragments predicted theoretically (Pfenniger \& Combes 1994). 
 These structures are now observed in a large number directly,
through VLBI in the vicinity of the Sun, through HI absorption in front of
quasars (e.g. Diamond e tal. 1989,
Davis et al 1996, Faison et al 1998). If this 10-20 AU
size is adopted for the low cut-off scale of the fractal structure,
the latter ranges over 6 orders of
magnitude in size, and about 10 in masses.

\bigskip
\subsection{ Scaling Laws}

\begin{figure}[t]
\psfig{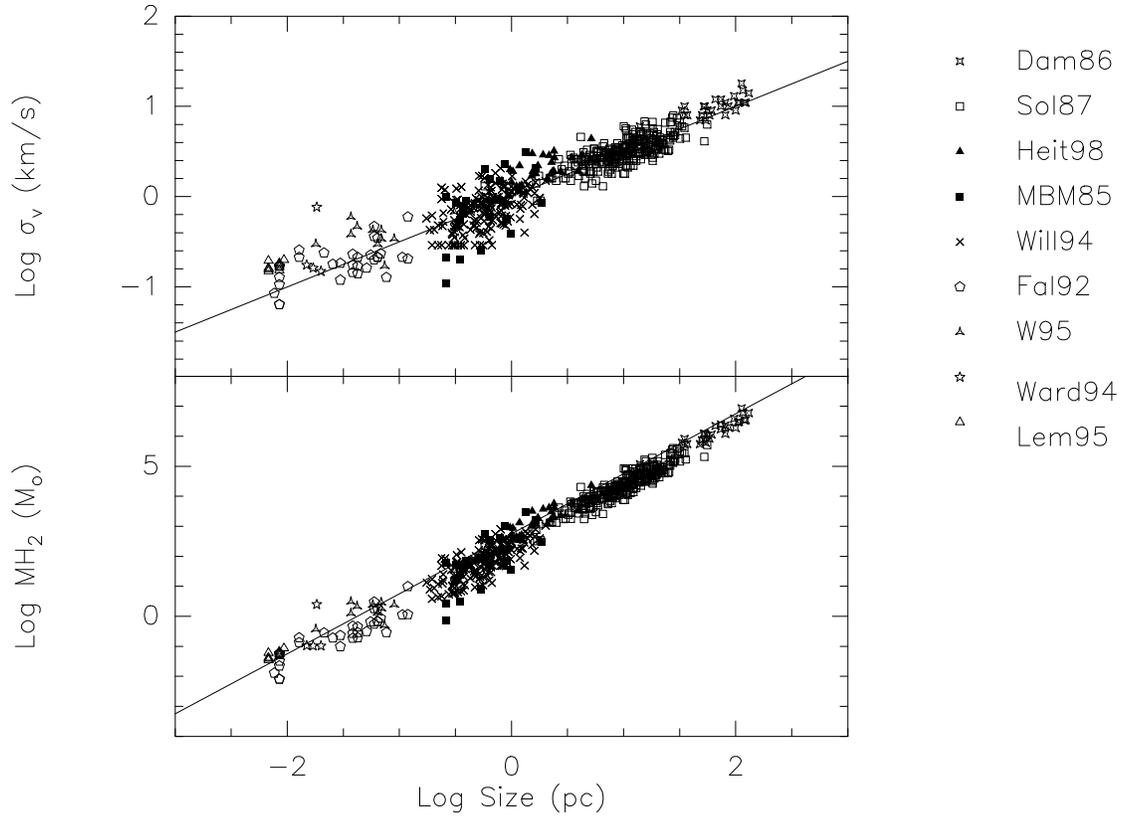}
\caption{{\it Top:} Size-linewidth relation taken from various sources:
Dam86:  Dame et al. (1986);
Sol87: Solomon et al. (1987);  
Heit98: Heithausen et al. (1998); 
MBM85: Magnani et al. (1985); 
Will94: Williams et al. (1994);
Fal92: Falgarone et al. (1992);
W95: Wang et al. (1995);
Ward94: Ward-Thompson et al. (1994);
Lem95: Lemme et al. (1995). An indicative line of slope 0.5 is drawn.
{\it Bottom:} Mass-size relation deduced from the previous one, 
assuming that the structures are virialised. The line drawn has
a slope of 2. }
\label{larson}
\end{figure}

To better quantify the self-similar structure, several works have
revealed that the interstellar clouds (either molecular or atomic)
obey power-law relations between size, linewidth and mass
(cf Larson 1981). These scaling relations are observed
whatever the tracer. The original one is the size-linewidth
relation (directly derived quantities), while the mass is
only a secondary quantity, very uncertain to obtain,
since there is no good universal tracer. The H$_2$ molecule does not
radiate in the cold conditions of the bulk of the ISM (10-15 K),
since it is symmetric, with no dipole moment. The first tracer is
the most abundant molecule CO (10$^{-4}$ 
in number with respect to H$_2$), but
it is most of the time optically thick, or photo-dissociated.
 The relation between the sizes $R$ and the line-widths 
or velocity dispersion $\sigma$, can be expressed through the power-law:

$$
 \sigma \propto R^q
$$
with $q$ between 0.3 and 0.5 (e.g. Larson 1981, Scalo 1985, 
Solomon et al 1987, cf Fig. \ref{larson}).
There are some hints that molecular clouds are virialised (at least at
large scale, since the masses are even more uncertain
at low scales and high densities). If the virial is assumed
at all scales, then:

$$
 \sigma^2 \propto M/R
$$
and the size-mass relation follows:

$$
M \propto R^D
$$
with $D$ the Hausdorff fractal dimension between 1.6 and 2.
It can be deduced also that the mean density over a given scale R
decreases as 1/R$^\alpha$, where $\alpha$ is between 1 and 1.4.

Other interpretations are possible (see e.g. Falgarone 1998).
Recently, Heithausen et al (1998) have extended the size-linewidth 
and size-mass relations down to Jupiter masses; their mass-size
relation is M$\propto r^{2.31}$, much steeper than previous
studies, but the estimation of masses at small scales is quite
uncertain (in particular the conversion factor between CO and H$_2$
mass could be higher, and the mass at small-scale underestimated).
It appears also that the self-similarity of structures is broken
in regions of star formation. The break is observed as a
change of slope in mass spectrum power laws at about 
0.05pc in scale in the Taurus cloud, corresponding to
the local Jeans length (Larson 1995). In other regions,
either this break does not occur, or is occuring at larger
scales (Blitz \& Williams 1997, Goodman et al. 1998).

Dense gas is molecular, and cold H$_2$ molecules
are difficult to trace (Combes \& Pfenniger 1997). 
The tracer molecules such as CO are either 
optically thick, or not thermally excited (in low-density regions),
photo-dissociated near ionizing stars, or depleted onto grains. 
Isotopic species, like
$^{13}$CO or C$^{18}$O, are poor tracers also, because of
selective photo-dissociation. The large range of scales
is also a source of bias, and of overestimates of the fractal
dimension: the small scales are not resolved, and 
observed maps are smoothed out. This process, which makes the
fractal look as a more diffuse medium of larger fractal dimension, 
can also lead to underestimates of the mass by factors more 
than 10 (e.g. simulations in Pfenniger \& Combes 1994).

The observations of molecular clouds reveal that the structure
is highly hierarchical, smaller clumps being embedded within the 
larger ones. This structure must be reminiscent of the formation 
mechanism, through recursive Jeans instability for instance.
Since we have no real 3D picture, it is however difficult
to ascertain a complete hierarchy, or to determine the importance
of isolated clumps and/or a diffuse intercloud medium.
An indicator of the 3rd dimension is 
the observed radial velocity, which is turbulent without
systematic pattern. It has been possible, however, to build a tree structure
where each clump has a parent for instance for clouds in Taurus 
(Houlahan \& Scalo 1992).

\bigskip
\subsection{  2D-Projection of the Fractal}

The projection of a fractal of dimension $D$ may not be a 
fractal, but if it is one with dimension $D_p$
it is impossible a priori to deduce its fractal dimension, 
except that 

$D_p = D $ if $D \leq 2$

$D_p = 2 $ if $D \geq 2$
(Falconer 1990).

In 2D, the fractal dimension can be measured by computing 
the surface versus the perimeter of a given structure. 
This method has been used in observed 2D maps, like the IRAS continuum
flux, or the extinctions maps of the sky. In all cases, this
method converge towards the same fractal dimension $D_2$.
For a curve of fractal dimension $D_2$ in a plane, the perimeter P and area
A are related by
$$
 P \propto A^{D_2/2}
$$
Falgarone et al (1991) find a dimension $D_2$ = 1.36 for CO contours
both at very large (degrees) and very small scales (arcmin), and the same
is found for IRAS 100$\mu$ contours in many circumstances (e.g. Bazell 
\& Desert 1988). Comparable dimensions ($D_2$ between 1.3 and 1.5) are
found with any tracer, for instance HI clouds
(Vogelaar \& Wakker 1994).

\bigskip
\section{  Turbulence }

\bigskip
\subsection{  Theory}

The interstellar medium is highly turbulent.
To quantify, we can try to estimate the
Reynolds number $R_e$, that separates the
laminar (low $R_e$) from turbulent regimes: 
$$
R_e = v l /\nu 
$$
where $v$ is the velocity, $l$ a typical dimension, and $\nu$ the
kinematic viscosity. Turbulent regimes are characterized by 
$R_e >> 10^3$, implying that the advection term v . $\nabla$ v
dominates the viscous term in the fluid equation. 
The turbulent state is characterised by
unpredictable fluctuations in density and pressure, and a cascade
of whirls. 
In the ISM, the viscosity can be estimated from the product of the
macroturbulent velocity (or dispersion) and the mean-free-path of cloud-cloud
collisions (since the molecular viscosity is negligible).
 But then the Reynolds number is huge ($\approx 10^9$), and the presence
of turbulence is not a surprise. 

This fact has encouraged many interpretations of the ISM structure
in terms of what we know from incompressible turbulence. 
In particular, the Larson relations have been found as a sign of the
Kolmogorov cascade (Kolmogorov 1941). In this picture, energy is dissipated
into heat only at the lower scales, while it is injected only at large
scale, and transferred all along the hierarchy of scales. Writing that 
the energy transfer rate $v^2/(r/v)$ is constant gives the relation
$$
v \propto r^{1/3}
$$
which is close to the observed scaling law, at least for the 
smallest cores (Myers 1983).
  The source of energy at large scale could be the differential
galactic rotation and shear (Fleck 1981).  This idealized view has 
been debated (e.g. Scalo 1987): it is not obvious 
that the energy cascades down without
any dissipation in route (or injection), given the large-scale shocks, flows, 
winds, etc... observed in the ISM. The medium is highly supersonic
(with Mach numbers larger then 10 in general), and therefore
very dissipative. Energy could be provided at intermediate
scale by stellar formation (bipolar flows, stellar winds,
ionization fronts, supernovae..).
Also, the interstellar medium is highly 
compressible, and its behaviour could be quite different from
ordinary liquids in laboratory.
However a modified notion of cascade could still be
applied, leading to a Burgers spectrum, with 
$$
v \propto r^{1/2}
$$
(see e.g. Vazquez-Semadeni et al. 1999).

The role of the magnetic field is still unclear. The intensity of
the field has been measured through Zeeman line effects
to be around a few $\mu$G in dense clouds (Troland et al. 1996,
Crutcher et al 1999), and the observations are compatible
with the hypothesis of equipartition between magnetic and
kinetic energy. The magnetic field therefore plays an important
role in energy exchange between the various degrees of freedom,
but cannot prevent gravitational instabilities and cloud collapse,
parallel to the field lines; although Gammie \& Ostriker (1996)
had found that magnetic waves could indeed provide some 
support in low dimensionality, in more realistic simulations
there is almost no difference between $B=0$ and 
strongly magnetized models, as far as the dissipation
and energy decay rates are concerned (Stone et al. 1998).

Besides, many features of ordinary turbulence are present in the ISM.
For instance, Falgarone et al (1991) have pointed out that the
existence of non-gaussian wings in molecular line profiles might be
the signature of the intermittency of the velocity field in turbulent
flows. More precisely, the $^{13}$CO average velocity profiles
have often nearly exponential tails, as shown by the velocity derivatives
in experiments of incompressible turbulence (Miesch \& Scalo 1995).
Comparisons with simulations of compressible gas give similar results
(Falgarone et al 1994).
Also the curves obtained through 2D slicing of turbulent flows
have the same fractal properties as the 2D projected images of the ISM;
their fractal dimension D$_2$ obtained from the perimeter-area relation
is also 1.36 (Sreenivasan \& M\'eneveau 1986).

More essential, the ISM is governed by strong fluctuations in density
and velocity. It appears chaotic, since it obeys highly non-linear
hydrodynamic equations, and there is coupling of phenomena at all scales.
 This is also related to the sensibility to initial conditions that
defines a chaotic system. The chaos is not synonymous of random
disorder, there is a remarquale ordering, which is reflected in the
scaling laws. The self-similarity over several orders of magnitude
in scale and mass means also that the correlation functions behave
as power-laws, and that there is no finite correlation length.
 This characterizes critical media, experiencing a second order
phase transition for example.

\bigskip
\subsection{ Turbulence Simulations }

A large number of hydrodynamical simulations have been run, in order
to reproduce the hierachical density structure of the interstellar
medium. However, these are not yet conclusive, since the dynamical
range available is still restricted, due to huge computational requirements.
 To gain in spatial resolution, 2D or even less (because of symmetries)
computations are performed, but often the results cannot be
generalised to 3D.

It has been argued that self-similar statistics alone can generate the
observed structure of the ISM, in pressureless turbulent flows without 
self-gravity (Vazquez-Semadeni 1994); however, only three levels of
hierarchical nesting can be traced.
 When heating and cooling processes are included, and since the
corresponding time-scales are faster than the dynamical tine-scales,
the gas can be derived by a polytropic equation of state, as
$ P \propto \rho^\gamma$, with $\gamma$ being the effective
polytropic index. The isothermal value $\gamma$ = 1 is
one of the possibilities, between 0 and 2 found in simulations.

From the size-linewidth relation $ \sigma \propto R^{1/2}$, and the second
observed scaling law $ \rho \propto R^{-1}$, it can be deduced that

$$
\sigma \propto \rho^{-1/2}
$$
and therefore, if the turbulent pressure $P$ is defined as usual by

$$ dP /d\rho = \sigma^2 $$
it follows that
$$
P \propto log\rho
$$
which is the logatropic equation of state, or "logatrope".
This behaviour has been tested in simulations (e.g. Vazquez-Semadeni et al
1998), but the logatrope has not been found adequate to represent 
dynamical processes occuring in the ISM (either hydro, or magnetic 2D
simulations). The equation of state of the gas would be more similar
to a polytrope of index $\gamma \approx 2$. But the results could
depend whether the clouds are in approximate equilibrium or not
(cf McLaughlin \& Pudritz 1996).

Vazquez-Semadeni et al (1997) have searched for Larson relations in
the results of 2D self-gravitating hydro (and MHD) simulations
of turbulent ISM: they do not find clear relations, but instead a
large range of sizes at a given density, and a large range of column
densities; the Larson relation is more the upper envelope
of the region occupied by simulated points in the 
$\rho$ - R diagram. They suggest that the observational results could be
artefacts or selection effects (existence of a threshold in
column density for UV-shielding for example).

MHD simulations have also explored the correlations between 
magnetic field strength and density, and field directions and
filamentary morphology of clouds. Again, there is no clear 
correlation, only a larger scatter, with an upper envelope
such as $B$ varying as $\rho^{0.4}$ (Padoan \& Nordlund
1998, Padoan et al. 1998).

Note also that chemical reactions network, combined with
turbulence, can be the source of 
considerable chaos (Rousseau et al. 1998).

\bigskip
\section{  Self-Gravity }

\bigskip
\subsection{  Theory}

Although the ISM is a self-organizing, multi-scale medium, comparable
to what is found in laboratory turbulence, there are very special 
particularities that are not seen but in astrophysics. Self-gravity
is a dominant, while it has not to be considered in atmospheric clouds
for instance. It has been recognized by Larson (1981) and by many others
that at each scale the kinetic energy associated with the linewidths 
balances the gravitational energy: clouds are virialized approximately,
given their very irregular geometry.

Self-gravitating gas in an isothermal regime (which is a 
good approximation for the ISM), is known to be subject to
instabilities. If we consider a sphere of gas confined 
in a box, and in contact with a thermostat, 
it will tend to follow an isothermal sphere, if the gas is hot enough.
Below a certain critical temperature, there is no equilibrium
any more, and the gas heats up when being cooled down,
it is the gravothermal catastrophy, caused by negative specific
heat (Lynden-Bell \& Wood 1968). Small sub-condensations
could form, and the physics will be more complex, since
in the asymptotic case of an isolated clump, it will evaporate
in a large number of dynamical times. The environment is
quite important, and the fact that the clumps can exchange
mass and energy with surroundings as well (e.g. Padmanabhan 1990). 

Gravitational instability and cloud collapse
is accompanied by fragmentation in a system
with very efficient cooling, and this process can provide the
turbulent motions observed. The theory was first proposed by
Hoyle (1953) who showed that the isothermal collapse of a cloud
led to recursive fragmentation, since the Jeans length decreases
faster than the cloud radius. Rees (1976) has determined the size of
the smallest fragments, when they become opaque to their own 
radiation. They correspond roughly to the smallest scales observed
in the ISM (sizes of 10 AU, and masses of 10$^{-3}$ M$_\odot$,
see the physical parameters of the "clumpuscules" in 
Pfenniger \& Combes 1994).

\begin{figure}[t]
\psfig{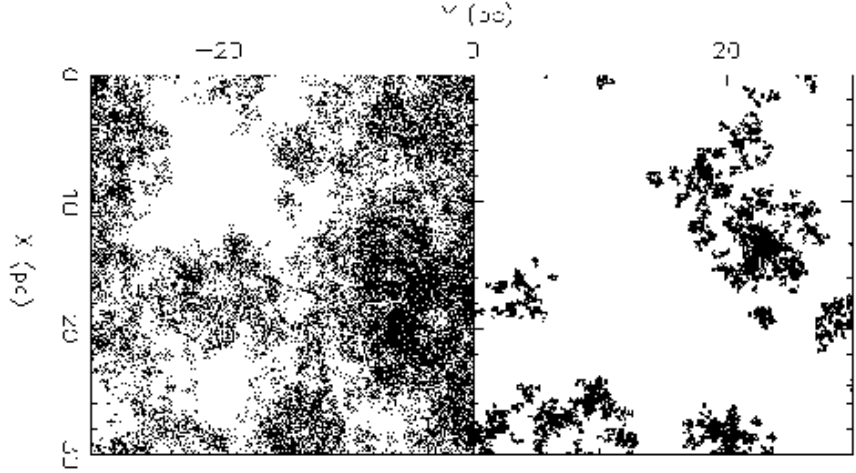}
\caption{{\it Left:} Simulations of a fractal of dimension $D=2.2$
built by recursive fragmentation;
{\it Right:} Same for  a fractal of dimension $D=1.6$,
the number of points are similar in the two cases.}
\label{fract}
\end{figure}

The hierarchical fractal structure yielded by recursive
fragmentation can be simulated schematically, in order to 
estimate the resulting filling factor, and the biases
introduced by observing with a limited spatial resolution
(e.g. Pfenniger \& Combes 1994). Clouds are distributed
according to the radial distribution of an isothermal
sphere in r$^{-2}$, and fragments also have positions
selected randomly according to this radial distribution,
and so on, hierarchically. The number $N$ of fragments at
each level, and the fractal dimension $D$ chosen suffice to determine
the size ratio between to imbricated spheres, i.e. = $N^{1/D}$. 
A sample of these fractals is shown in Fig. \ref{fract}, for
two dimensions 2.2 and 1.6.

Many other physical processes play a role in the turbulent ISM, as
for instance rotation and magnetic fields. But they cannot be identified as
the motor and the origin of the structure. Galactic rotation certainly
injects energy at the largest scales, but angular momentum cannot
cascade down the hierachy of clouds; indeed if the rotational velocity
is too high, the structure is unstable to clump formation (cf Toomre
criterium, 1964), and the non-axisymmetry evacuates angular momentum outside
the structure. Magnetic fields are certainly enhanced by the turbulent
motions, and could reach a certain degree of global equipartition
with gravitational and kinetic energies in tbe virialised clouds.
 But they cannot be alone at the origin of the hierarchical structure,
gravity has to trigger the collapse first. Besides, there is
no observational evidence of the gas collapse along the field lines,
polarisation measurements give contradictory results for the field
orientation with respect to the gas filaments. Therefore, although
rotation, turbulence, magnetic fields play an important role in the ISM,
they are more likely to be consequences of the formation of the structure.

\bigskip
\subsection{ Self-gravitating Simulations}
  
Simulations of self-gravitating gas are very demanding,
since large gradients rapidly set up, and spatial resolution
must be adapted. A general rule is that the resolution
is well below the Jeans length at any point, but the Jeans
length shrinks along the collapse. Artificial fragmentation 
can sometimes happen due to artifacts (see e.g. Truelove et al. 1997).
To compensate for the limited spatial range, periodic boundary
conditions are used, to simulate the ISM.
Klessen (1997) and Klessen et al. (1998) have considered the
fragmentation of molecular clouds, in 3D, starting with 
an homogenous cloud with small primordial fluctuations.
 These initial conditions are very similar to what is used
in a cosmological background. The fluctuations are a
gaussian density field with power spectrum P(k) $\propto$ k$^{-2}$.
After one free-fall time, the gas has evolved into a system 
of filaments and knots, some of them contain collapsing cores.
 To avoid the problem of spatial resolution, the condensed
cores are then replaced by sink particles, simulating
therefore a low cut-off scale.

Klessen et al. (1998) compute the mass spectrum of clumps, which
looks very similar to the observed one $dN/dm \propto m^{-1.5}$
in the ISM, at least over 1.5 order of magnitude.
The results however, depends still a lot on the
initial conditions, and the power spectrum used
(Semelin \& Combes 1999). 
 One of the problem of these simulations, in comparison
with the ISM, is the absence of energy re-injection, and 
systematic motions. In a Galaxy, differential rotation
and shear should provide both.

\bigskip
{\it Taking into Account the Shear}

One way to re-inject energy at large scale is the differential
rotation of the Galaxy, as we have already emphasized. 
Computations of self-gravitating shearing sheets have
already given interesting large-scale structures, that
might be similar ro fractals. 
Toomre (1990) carried out self-gravitating simulations
on gas particles, that were regularly cooled, in a shearing
environment, and a quasi-periodical boundary conditions.
 In fact, the periodicity has to take into account the
shear motions, and proper sliding of the sheet at larger 
and smaller radii is necessary in order to ensure continuity.
The main result is a steady wave pattern, that sets quickly 
in through gravitational instabilities, and differential
rotation, and is continuously renewed. Toomre (1990) 
was surprised himself of the efficiency of the mechanism,
and of its quasi-stationarity: the energy dissipated in
the gas cooling was compensated exactly by the rotational
shear. This could be the main energy source was the
gravitational structures in the ISM of galaxies.
Toomre \& Kalnajs (1994) refined the computations, and
obtained the "typical" morphology of the pattern.
These simulations demonstrate the power of gravity,
allied with shear, to produce spiral structure,
by the swing mechanism. They also show how much structures
at all scales, or spiral chaos, can be sustained and maintained
by the self-gravitation of a cooled granular distribution.

The work was taken up recently by Huber \& Pfenniger
(1999), that generalised the computations in 3D, 
taking into account the galaxy plane thickness of the 
gaseous component. Their cooling is simulated by a viscous
force, proportional to the particle velocity. They 
also find the characteristic filaments due to the
shear combined with self-gravity, and measure the
corresponding fractal dimension of the structures.
 Since the filaments are much longer in the plane 
than thick (perpendicular to it), the fractal 
dimension is in fact dominated by the z-dimension
plane morphology, at least at large scales.
 So the range of scale where the fractal dimension 
is lower than 2 is very small. The boarder effects
are then too large to determine $D$ without ambiguity.
These calculations are however encouraging that
a fractal structure could develop at sub-kpc scales,
when shearing is dominant, near the high cut-off
of the ISM fractal.

\begin{figure}[t]
\psfig{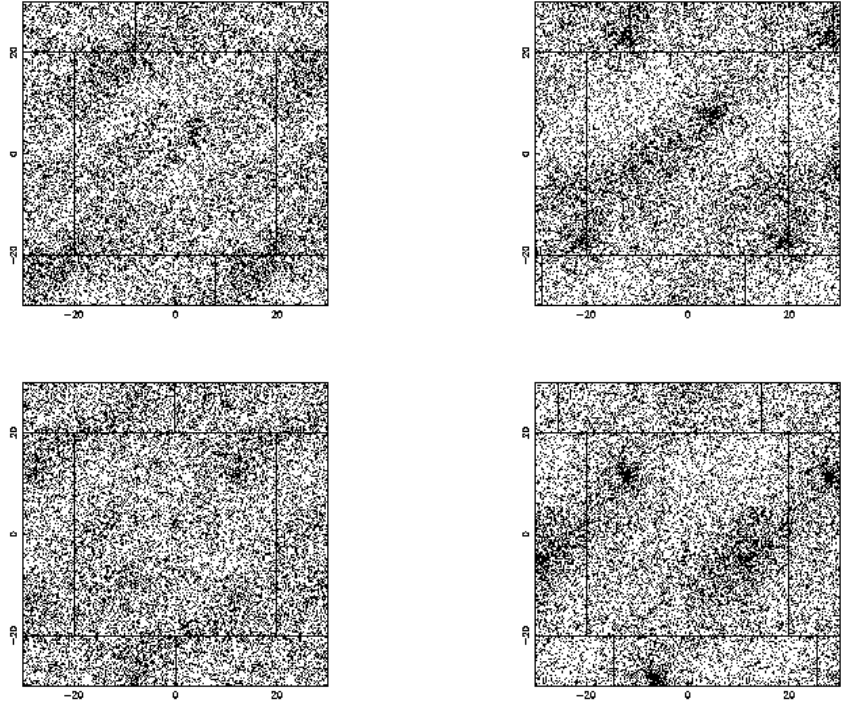}
\caption{ {\it From Left to Right, and Top to Bottom:} Simulations in 2D of
a shearing sheet of gas particles, subject to their own self-gravity,
and to collisions (dissipation). The central square in each frame is the 
area simulated, and the lanes surrounding it have been plotted
to show the particular periodicity, that takes into account the shear
(relative X-motions). The gradient of the rotational angular velocity
$\Omega$ is along the Y-axis.
The time is 1, 4, 11, and 16 free-fall times respectively
for the four successive frames.}
\label{shearing}
\end{figure}

In a recent work (Semelin \& Combes, 1999), we
have also tried to determine the fractal dimension
of a nearly isothermal self-gravitating gas. A 
tree-code and cloud-cloud collisions algorithm was used 
in 2D and 3D to follow the collapse of a periodically-replicated
piece of ISM. Without shear, the collapse
of initial perturbations are followed , and a fractal structure
is found only in a transient way. With energy re-injection,
and in particular with shear (and Coriolis forces),
the structures are formed over several orders of
magnitude. The characteristic spiral filaments are
observed, and embedded clumps form, and can
disappear and reform stochastically.
Simulations of a shearing sheet are shown in Fig \ref{shearing}.
 The fractal dimension derived is around 1.8.

Wada \& Norman (1999) have also simulated a shearing 
galaxy disk, but with a multi-phase medium: they obtain 
the formation of dense clumps and filaments, surrounded
by a hot diffuse medium. The quasi-stationary filamentary
structure of the cold gas is another manifestation of
the combination of shear and self-gravity in a cooled gas. 

\bigskip
\section{  Galaxy Distributions}

It has been known for a long time that galaxies are not distributed
homogeneously in the sky, but they follow a hierarchical 
structure: galaxies gather in groups, that are embedded in
clusters, then in superclusters, and so on (Charlier 1908, 1922,
Shapley 1934, Abell 1958). 
Moreover, galaxies and clusters appear to obey scaling properties,
such as the power-law of the two point-correlation function:
$$
\xi(r) \propto r^{-\gamma}
$$
with the slope $\gamma$, the same for galaxies and clusters, of $\approx$ 1.7
(e.g. Peebles, 1980, 1993). 
According to the self-similar morphology, and the scaling laws,
the galaxy ensemble can also be characterized by a fractal.
But what are the limiting scales?

\bigskip
\subsection{ Low and High-mass cut-off for the galaxies fractal}

The smallest structure is of course a galaxy, and the starting
point of the fractal is therefore obvious: 10 kpc in scale,
and about 10$^{10}$ M$_\odot$, representative of a dwarf
galaxy. The high-mass cut-off is much less obvious, and 
there has been a continuous debate in the recent years
to know what is the scale of transition to homogeneity,
and even whether this scale exists.

Isotropy and homogeneity are expected at very large scales from the
Cosmological Principle (e.g. Peebles 1993). 
The main observational evidence in favor of the Cosmological Principle
is the remarkable isotropy of the
cosmic background radiation (e.g. Smoot et al 1992), that provides information
about the Universe at the matter/radiation decoupling. At very large scales,
the Universe must then be homogeneous. There must 
exist a transition between the small-scale fractality to large-scale
homogeneity. This transition is certainly smooth, and might correspond to the
transition from linear perturbations to the non-linear gravitational collapse 
of structures. The present catalogs do not yet see the transition since
they do not look up sufficiently back in time. It can be noticed that
some recent surveys begin to see a different power-law
behavior at large scales ($\lambda \approx 200-400  h^{-1}$ Mpc, e.g. 
Lin et al 1996, Scaramella et al. 1998). The interpretation depends
however on the K-correction adopted, and the curvature of the Universe
(Joyce et al. 1999).

Summarizing, it appears that the high-mass cut-off could be at
the present epoch at about 300 Mpc and 10$^{17}$ M$_\odot$,
the mass of the largest superclusters. The fractal extends over
about 4 orders of magnitude is scale and 7 in mass. It is smaller
than the ISM fractal, but since larger and larger structures could
decouple from inflation and develop, this range could still
increase.

\bigskip
\subsection{  Correlation Function and Conditional Density }

The debate on the spatial extent of the fractal has been 
complexified by the use of the correlation function to
quantify the scaling laws in galaxy distributions. 
The correlation function is defined as
$$
\xi(r) = \frac{<n(r_i).n(r_i+r)>}{<n>^2} -1
$$
where $n(r)$ is the number density of galaxies, and $<...>$ is the volume
average (over $d^3r_i$). One can always define a 
correlation length $r_0$ by $\xi(r_0) = 1$.

This definition involves the average density $<n>$, which depends on
the scale for a fractal distribution. This is unfortunate, since 
the derived correlation parameters (slope and correlation 
lengths) then depend on the galaxy sample used 
(see Davis 1997, Davis et al 1988, Pietronero et al 1997). 
 The fact that there exists a correlation length does not mean 
that there is no fractal, because of its definition different from 
that used in physics; there $\xi_0$ characterizes the exponential decay of 
correlations $ (\sim e^{- r/ \xi_0} ) $ (for power decaying correlations, the
correlation length is infinite).
Davis \& Peebles (1983) or Hamilton (1993) argue that
the fractal of galaxies cannot have a large spatial extent, since
the galaxy-galaxy correlation length $r_0$ is rather small. 
The most frequently reported value is $r_0 \approx 5 h^{-1}$ Mpc
(where $h = H_0$/100km s$^{-1}$Mpc$^{-1}$).

The same problem occurs for the two-point correlation function of
galaxy clusters; the corresponding $\xi(r)$ has the same power law 
as galaxies, their  length  $r_0$ has been reported to be about 
$r_0 \approx 25 h^{-1}$ Mpc, and their correlation amplitude is therefore
about 15 times higher than that of galaxies
(Postman, Geller \& Huchra 1986, Postman, Huchra \& Geller 1992).
The latter is difficult to understand, unless there is a considerable
difference between galaxies belonging to clusters and field galaxies (or
morphological segregation). The other obvious explanation is that
the normalizing average density of the universe was then chosen lower.

Assuming that the average density is a constant, while homogeneity
is not yet reached, could perturb significantly the correlation 
function, and its slope, as shown by Coleman, Pietronero \& Sanders (1988) 
and Coleman \& Pietronero (1992).
 The function $\xi(r)$ has a power-law behaviour 
of slope $-\gamma$ for $r< r_0$, then it turns down to zero 
rather quickly at the statitistical limit of the sample. This rapid
fall leads to an over-estimate of the small-scale $\gamma$.
Pietronero (1987) introduces the conditional density
$$
\Gamma(r) = \frac{<n(r_i).n(r_i+r)>}{<n>} 
$$
which is the average density around an occupied point.
For a fractal medium, where the mass depends on the size as
$M(r) \propto r^D$, 
$D$ being the fractal (Haussdorf) dimension, the conditional
density behaves as  $\Gamma(r) \propto r^{D-3}$.

\begin{figure}[t]
\psfig{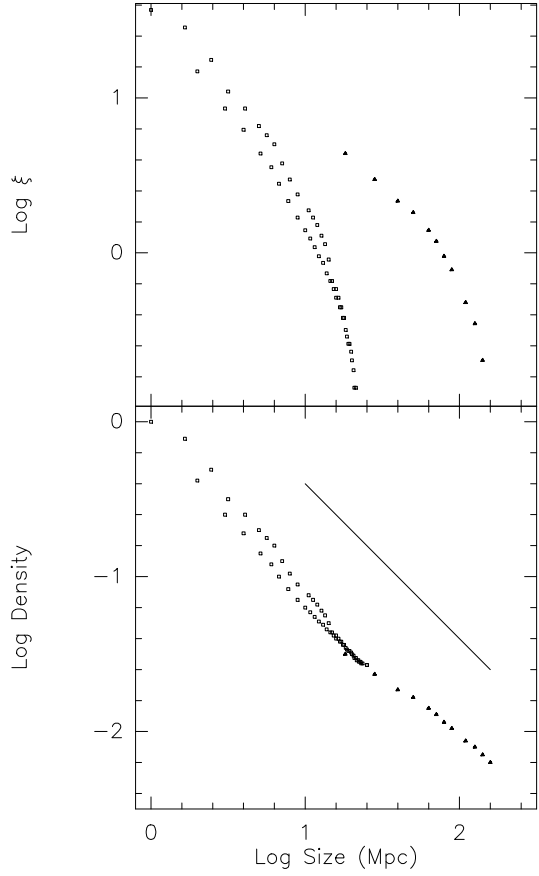}
\caption{ {\it Bottom:} the average conditional density $\Gamma(r)$ for
several samples (Perseus-Pisces and CfA1 in open rectangles, and LEDA in
filled triangles, adapted from Sylos-Labini \& Pietronero 1996).
{\it Top:} $\xi(r)$ corresponding to the same surveys.
The indicative line has a slope $\gamma=1$ (corresponding to a fractal dimension
$D\sim2$). This shows that it is difficult to determine the slope on
the $\xi(r)$ function.}
\label{galdib}
\end{figure}

It is possible to retrieve the correlation function as
$$
\xi(r) = \frac{\Gamma(r)}{<n>} -1
$$
In the general use of
$\xi(r)$,  $<n>$ is taken for a constant, and we can see that
$$
D = 3 - \gamma \quad .
$$
If for very small scales,
both $\xi(r)$ and $\Gamma(r)$ have the same power-law behaviour, with the 
same slope $-\gamma$, then the slope appears to steepen for $\xi(r)$
when approaching the  length $r_0$. This explains why
with a correct statistical 
analysis (Di Nella et al 1996, Sylos Labini \& Amendola 1996, 
Sylos Labini et al 1996), the actual $\gamma \approx 1-1.5$ is smaller 
than that obtained using $\xi(r)$ (cf Fig. \ref{galdib}).
 This also explains why the amplitude of
$\xi(r)$ and $r_0$ increases with the sample size, and for clusters as well. 

Note that the fractal distribution in the galaxy catalogs has
been determined from the light distribution, and this
could be somewhat different from the mass distribution.
There is also a morphological  segregation of galaxies 
in clusters, and the different types of galaxies (ellipticals,
spirals, or dwarfs) do not trace the same distribution.
The mass distribution could be more complex.
All this has led to the introduction of multifractality
to represent the Universe (e.g. Sylos-Labini \& Pietronero 1996).
In a multifractal system, local scaling properties slightly
evolve, and can be defined by a continuous distribution of exponents.
This is a mere generalisation of a simple fractal, that links the space
and mass distributions. 
Mutlifractality may also better account for the transition to homogeneity,
with a fractal dimension varying with scale
(Balian \& Schaeffer 1989, Castagnoli \& Provenzale 1991, 
Martinez et al 1993, Dubrulle \& Lachi\`eze-Rey 1994).

\bigskip
\subsection{Definition of the Self-Gravity Domain}

The general view in cosmologiacl models is that the galaxy structures in the 
Universe have developped by gravitational collapse from primordial fluctuations.
Once unstable, density fluctuations do not grow as fast as we are used
to for Jeans instability (exponential), since they are slowed down
by expansion. The rate of growth is instead a power-law, in the
linear regime. If   $\delta$ is the density contrast:
$$ \delta(\vec x) = (\rho(\vec x) -<\rho>)/<\rho> $$
where $<\rho>$ is the mean density of the Universe, assumed
homogeneous at very large scale.
If $\vec r$ is the physical coordinate, the comoving coordinate 
$\vec x$ is defined by
$ \vec r = a(t) \vec x $,
where a(t) is the scale factor, accounting for the Hubble expansion
(normalised to a(t$_0$) = 1 at the present time). Since the Hubble constant
verifies $H(t) = \dot{a}/a$, the peculiar velocity is defined by
$$ \vec v = \dot{\vec r} - H \vec r = a \dot{\vec x}$$
In comoving coordinates, the Poisson equation becomes:
$$ \nabla_x^2\Phi = 4\pi G a^2 (\rho -<\rho>)$$
 It can be shown easily that in a flat universe, the density contrast
in the linear regime grows as the scale factor $a(t) = (t/t_0)^{2/3}$.
In the non-linear regime, on the contrary, the density $\rho$ is
much larger with respect to $<\rho>$, and the normal self-gravity
and Jeans instability is retrieved. The receding velocities due to 
inflation do not exist anymore. The structures are bound and decoupled
from inflation. It is natural to define the self-gravity domain
as limited by the largest decoupled structures, and if self-gravity
is responsible for the fractal structure, it is expected that this scale 
will also delimit the transition to homogeneity.

\bigskip
\section{Statistical Mechanics of Self-Gravity}

Both for the ISM and for galaxy distributions in the Universe,
self-similar structures are observed over large ranges in scales.
 Scaling laws are observed, which translate by an average density
decreasing with scale as a power-law, of slope $-\gamma$ between -1.5 and -1,
corresponding to a fractal dimension $D = 3-\gamma$ between 1.5 and 2.
In the following, we will investigate the hypothesis that the 
main driver responsible for these fractal structures is
self-gravity, even though the actual structures migh be more
complex and perturbed.

Since gravity is scale-independent, there are opportunities for 
a mechanism to propagate over scales in a self-similar fashion.
For the ISM, in a quasi isothermal regime, a fractal structure
could be build through recursive Jeans instability and fragmentation. 
 This recursive fragmentation proceeds until the density is high enough to
reach the adiabatic regime. Self-gravity could be the principal 
origin of the fractal, with generated turbulent motions in virial
equilibrium at each scale. For galaxy formation, the smallest 
structures collapse first, and these influence the largest scale
in a non-linear manner. It is obvious that in both cases, the system 
does not tend to a stationary point, but develops fluctuations
at all scales, and these must be studied statistically.

Recently, de Vega, Sanchez \& Combes (1996a,b, 1998)
have proposed a statistical field theory of self-gravity,
not only to account for the existence of the structure, but also
to be able to predict its fractal dimension and others critical exponents.
This has been obtained by developping
 the grand partition function of the ensemble of
self-gravitating particles. In transforming
the partition function through a functional integral, 
it can be shown that 
the system is exactly equivalent to a scalar field theory. The theory
does not diverge, since the system is considered only between two
scale limits: the short-scale and large-scale cut-offs.
Through a perturbative approach it can be demonstrated that the system 
has a critical behaviour, for any parameter
(effective temperature and density). That is, we can consider the
self-gravitating gaseous medium as correlated at any scale, as for
the critical points phenomena in phase transitions 
(as was first suggested by  Totsuji \& Kihara 1969).

Note that this approach is quite different from the 
thermodynamical approach developped by 
Saslaw \& Hamilton (1984). Their theory is based on
the thermodynamics of gravitating systems,
which assumes quasi thermodynamic
equilibrium. They justified this equilibrium at the small-scales of non-linear
clustering, because the local relaxation and dynamical 
time-scales are much shorter than the expansion time-scale. 
The theory considers the essential parameter $b(t)$,
the ratio of gravitational correlation energy to thermal kinetic energy,
and deduces the value of this parameter from the observations.
It appears that $b(t)$ varies also with scale.
  The predictions of the thermodynamical theory have been successfully
compared with N-body simulations (Itoh et al. 1993, Sheth \& Saslaw 1996,
Saslaw \& Fang 1996).

\bigskip
\subsection{  Hamiltonian of the Self-Gravitating  Ensemble of N-bodies}

  Let us consider a gas of particles submitted only to their self-gravity,
in thermal equilibrium at temperature T  $(kT = \beta^{-1}) $.
In the interstellar medium, quasi isothermality is justified,
due to the very efficient cooling. For unperturbed gas in 
the outer parts of galaxies, gas is in equilibrium with the cosmic background
radiation at $T \approx 3K$ (Pfenniger et al 1994, Pfenniger \& Combes 1994).
For a system of collapsing structures in the universe, this
can be a valid approximation, as soon as the gradient of
temperature is small over a given scale.

This isothermal character is essential for the description of the
gravitational systems as critical systems, as will be shown
later, so that the canonical ensemble appears the best adapted system.
Moreover, the systems considered are not
isolated gravitational systems, for which the microcanonical
system should be used (e.g. Horwitz \& Katz 1978a,b; Padmanabhan 1990).
On the contrary, the mass or number of particles is not
fixed, and according to the fluctuations, matter can enter any
given scale, through condensation or evaporation.
The best statistical frame to consider is then
the grand canonical ensemble, allowing for a variable
number of particles $N$.
The grand partition function ${\cal Z}$ and the Hamiltonian $H_N$ are

$$
{\cal Z} = \sum_{N=0}^{\infty}\; {{z^N}\over{N!}}\; \int\ldots \int
\prod_{l=1}^N\;{{d^3p_l\, d^3q_l}\over{h^3}}\; e^{- \beta H_N}
$$

$$
H_N = \sum_{l=1}^N\;{{p_l^2}\over{2m}} - G \, m^2 \sum_{1\leq l < j\leq N}
{1 \over { |{\vec q}_l - {\vec q}_j|}}
$$
where $z$ is the fugacity = $exp(-\beta \mu_c)$ in terms of the
gravito-chemical potential $\mu_c$, and $h$ is now the Planck 
constant.

The latter integral can be transformed, using the continuous 
density
$\rho({\vec r})= \sum_{j=1}^N\; \delta({\vec r}- {\vec q}_j)\; $
and integrated to yield the potential, but introducing a 
 cutoff $ a $ for the minimum separation between particles,
so that there is no problem of divergence.
The cutoff $ a $ is here introduced naturally, it corresponds
to the size of the smallest fragments, or clumpuscules
(of the order of $\sim 10$ AU). In fact, we consider that the
particles of the system interact with the Newton
law of gravity ($1/r$) only within the size range of the fractal,
where self-gravity is predominant. At small scale, other forces
enter into account, and we can adopt a model of hard spheres
to schematize them. Also at large scales, beyond
the upper cutoff, different forces must be introduced.
The phenomenological potential thus considered does not possess
any singularity.

\bigskip
\subsection{  ${\cal Z}$ as the Partition Function of a Single Scalar Field $\phi$ }

Using the potential in $1/r$, and its inverse operator 
$-\frac{1}{4\pi}\nabla^2$
(but see also a similar derivation, with $[1 - \theta (a-r) ]/r$ 
and its corresponding inverse operator,
for the phenomenological potential with cutoff, in
de Vega et al 1996b), the exponent of the
potential energy can be represented 
as a functional integral (Stratonovich 1958, Hubbard 1959)
$$
e^{  \frac12\, \beta G \, m^2
\int \;
{{d^3x\, d^3y}\over { | {\vec x} - {\vec y}|}}\; \rho({\vec x})
\rho({\vec y})} = \int\int\; {\cal D}\xi \; e^{ -\frac12\int d^3x \; (\nabla
\xi)^2 \; + \; 2 m \sqrt{\pi G\beta}\; \int d^3x \; \xi({\vec x})\;
\rho({\vec x}) } 
$$

With the change of variables:
$\phi({\vec x}) \equiv  2 m \sqrt{\pi G\beta}\;\xi({\vec x}) $
and
$$
\mu^2 = {{\pi^{5/2}}\over {h^3}}\; z\; G \, (2m)^{7/2} \, \sqrt{kT} \; 
\quad , \quad T_{eff} = 4\pi \; {{G\; m^2}\over {kT}} \quad 
$$
the partition function can be written as a functional integral
${\cal Z} =  \int\int\;  {\cal D}\phi\;  e^{ -S}$
where the action $S$ is:
$$
S[\phi(.)] \equiv  {1\over{T_{eff}}}\;
\int d^3x \left[ \frac12(\nabla\phi)^2 \; - \mu^2 \; e^{\phi({\vec
x})}\right] 
$$

Note that the "equivalent" temperature $T_{eff}$ in the field theory
is in fact inversely proportional to the physical temperature.
It can be shown that the parameter $\mu$ is equal to the inverse
of the Jeans length, itself of the order of the cutoff $a$.

It is then possible to compute the statistical average  
of physical quantitites, such as the density
$\rho({\vec r})$ 

$$
  <\rho({\vec r})> =  -{1 \over {T_{eff}}}\;<\nabla^2 \phi({\vec r})>=
{{\mu^2}\over{T_{eff}}} \; <e^{\phi({\vec r})}>  
$$
where $ <\ldots > $ means functional average over   $ \phi(.) $
with statistical weight $  e^{S[\phi(.)]} $. 

The equation for stationary points:
$\nabla^2\phi = -\mu^2\,  e^{\phi({\vec x})} \; $
has two main solutions. One is the
constant stationary solution:
 $ \phi_0 = -\infty $; and the second is the singular 
isothermal sphere:
$\phi(r) = \log{{2}\over { \mu^2 r^2}} $.
With a perturbative method,
starting from the stationary solution $ \phi_0 = -\infty $,
it has been shown that the theory scales, at large distances
 (de Vega et al 1996b). The same has been shown,
starting from the isothermal sphere solution
(Semelin et al. 1998). In the latter case, the development
in series of the density correlation function is made
with the relevant effective coupling constant:
$$
\lambda = { {T_{eff}} \over {R} }
$$
and this coupling constant evolves through the
renormalisation group equations with the scale $R$, or
scale ratio $\tau = ln(R/a)$, where $a$ is
the low cut-off scale. A remarkable behaviour is found
for $\lambda(\tau)$, since it vanishes periodically, for
values $\tau_n$ = $2\pi n /\sqrt 7$ ($n$ integer).
Periodically, the coupling constant diverges to infinity, which
means that the perturbation theory is no longer valid,
because of strong coupling. This occurs for scales:
$$
R_n = R_0 e^{2\pi n /\sqrt 7} = R_0 (10.749)^n
$$ 
It appears then a hierachy of scale, varying
by about an order of magnitude at each level.
This numerical factor 10.749 depends essentially
on the spherical geometry assumed for the computations,
but is expected to be different for different geometries
(like filaments, for example).

\bigskip
\section{ Renormalization Group Methods}

The renormalization methods are very powerful to deal with
self-similar systems obeying scaling laws, like critical
phenomena. In the latter case, examplified by second 
order phase transitions, there exist critical divergences,
where physical quantities become singular as power-laws
of parameters called critical exponents.
These critical systems reveal a collective behaviour,
organized from microscopic degrees of freedom, through giant
fluctuations and statistical correlations.
Hierarchical structures are built up, coupling all scales
together, replacing an homogeneous system
in a scale-invariant system. It can be shown that 
local forces are not important to describe the collective behaviour,
which is only due to the statistical coupling of local
interactions. Therefore, critical exponents depend only
on the statistical distribution of microscopic
configurations, i.e. on the dimensionalities or
symmetries of the system. There exist wide universality
classes, that allow to draw quantitative predictions
on the system from only a qualitative knowledge of
its properties (e.g. Parisi, 1988; Zinn-Justin 1989; Binney et al 1992).

\bigskip
\subsection{  Critical Phenomena}

Critical phenomena occur at second order phase transitions,
i.e. continuous transitions without latent heat. The paradigm of
these systems is the transition at the Curie point (T=T$_c$=
1043K) from paramagnetic iron, where the magnetic moment 
is proportional to the applied field m=$\mu$ B, to ferromagnetic
state, where there exists a permanent magnetic moment m$_0$
even in zero field. 
Although the permanent magnet tends to zero continuously at T$_c$,
there are divergences: for instance the heat capacity C
behaves as C$ \propto | T - T_c |^{-\alpha}$, with $\alpha > 0$.

\begin{figure}[t]
\psfig{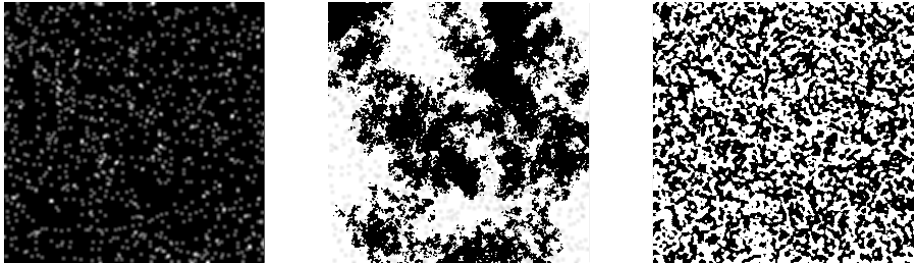}
\caption{ Illustration of the fractal structure occuring when a Ising system
passes through the critical temperature (a second order phase transition).
 The temperature is near zero at left,
about critical in the center, and much larger than critical at right
(from a simulation of the Ising model with a Monte-Carlo method) }
\label{tcrit}
\end{figure}

Also for the critical point of water, at which the transition from the liquid
to gas becomes continuous, the compressibility $\kappa_T \propto
 | T - T_c |^{-\gamma}$.
At the critical point, it is easy to understand that
the compressibility which tends to infinity generates
large density fluctuations, and therefore light is
strongly diffused by the varying optical index:
this is the critical opalescence. The extraordinary
fact is that microscopic forces can give rise
to large-scale fluctuations, as if the medium was
organized at all scales (cf Fig. \ref{tcrit}).

\subsection{  Universality of Critical Exponents}

Experiments have shown that the critical exponents
for a wide variety of systems are the same, and more
precisely they belong to universality classes, depending
only on the dimensionality $d$ of space and $D$ of
the order parameter (for instance if the field is scalar
or a vector with dimension $D$).
This universality means that the details of the local forces
are unimportant; therefore the local interactions
can be simply modelled, through a schematic hamiltonian
supposed to hold the relevant symmetries
of the system.
There exists only 2 independent critical exponents.

Renormalization techniques were 
applied to statistical physics in the 1970s
(Wilson \& Kogut 1974; Wilson 1975, 1983).
In a renormalization group transformation,
the scales are divided by a certain factor $k$, and
particles are replaced by a block of particles.
Since the system is scale-independent,
it should be possible to find an hamiltonian for the blocks
which is of the same structure as the original one.
The new system is less critical than the previous one, since 
the correlation length $\xi$ has also been divided by $k$.
It is a way to reduce the number of degrees of freedom 
of the system.
With these techniques, precise values of the critical
exponents have been computed, for a whole range
of models, and can be used for any problem
in the same universality class.

\bigskip
\subsection{  Statistical Self-Gravity }

As was shown in section 6, the grand-canonical self-gravitating
system is critical for a large range of the parameters, and
it is difficult to isolate a critical point, to identify diverging
behaviours. However, it is 
well known (Wilson 1975, Domb \& Green  1976), that
physical quantities diverge only for infinite volume systems,
at the critical point. Since the self-gravitating systems are also finite and
bounded, they only approach asymptotically the divergences.

If $ \Lambda $ measures the distance to the critical point,
 (in spin systems for instance, $ \Lambda $ is
proportional to $| T - T_c |$),
the correlation length  $ \xi $ diverges like 
$$ 
\xi( \Lambda ) \sim  \Lambda^{-\nu} \; 
$$
and the specific heat (per unit volume) as
$ {\cal C} \sim  \Lambda^{-\alpha}  \; $.
But in fact, for a finite volume system, all physical quantities are 
finite at the critical point.
When the typical size $R$ of the system 
is large, the  physical  magnitudes
take large values at the critical point, and
the infinite volume theory is used to treat finite size systems at
criticality. 
In particular, for our system, the correlation length provides the
relevant physical length $ \xi \sim R $, and we can write
$$
\Lambda \sim R^{-1/\nu} \; 
$$

The self-gravitating systems considered here 
 have the symmetries $d=3$ and $D=1$ (scalar field), which 
should indicate the universality class to which it corresponds.
It remains to identify the corresponding operators. Already in the previous
sections, it was suggested that the field $\phi$ corresponds
to the potential, and the mass density  
$$
m\, \rho({\vec x}) = m\, \,  e^{\phi({\vec x})}
$$
can be identified with the  energy density in the renormalization
group (also called the `thermal perturbation operator').

We note that the state of zero density (or zero fugacity),
corresponds to a singular point, around which we develop the
physical functions (and we choose $\Lambda$ accordingly).
At this point $ \mu^2/T_{eff} = 0 $, the partition function $ {\cal Z}
$ is singular
$$
  \Lambda \equiv   {{\mu^2}\over{T_{eff}}} = z\,
  \left({{2\pi mkT}\over{h^2}}\right)^{3/2} \; 
$$
i.e., the critical point $  \Lambda = 0 $ corresponds to zero
fugacity $z$. Writing $ {\cal Z} $ as a function of the 
action $S^*$ at the critical point
$$
{\cal Z}(\Lambda) = 
 \int\int\;  {\cal D}\phi\;  e^{ -S^* + \Lambda
\int d^3x  \; e^{\phi({\vec x})}\;}
$$
and computing statistical averages (de Vega et al. 1996a,b),
the mass fluctuations and corresponding dispersion can be found as:
$$
(\Delta M(R))^2 \equiv  \; <M^2> -<M>^2 \sim
\int d^3x\; d^3y\; C({\vec x},{\vec y}) \sim R^{2/\nu}
$$
$$
\Delta M(R)  \sim  R^{1/\nu}
$$
This is the definition relation of the fractal, with dimension $d_H$,
and the scaling exponent $\nu$ can be identified with the inverse
Haussdorf dimension of the system,
$d_H = \frac1{\nu} $.
The velocity dispersion follows: $\Delta v \sim R^{q} $,
with $q =\frac12\left(\frac1{\nu} -1\right) =\frac12(d_H -1)  $.

The scaling exponents $ \nu , \; \alpha $ have been
computed through the renormalization group approach. The case of a 
single component (scalar) field has been extensively studied 
in the literature (Hasenfratz \& Hasenfratz 1986, Morris 1994a,b). 
For the Ising model $d=3$, the exponent
$\nu$ = 0.631, from which we deduce $d_H$ = 1.585.
Alternatively, in the case of weak perturbations, the
mean field theory can be applied, and $d_H$ = 2.
These values are compatible with the observed ones
for astrophysical fractals.

\bigskip
\section{ Conclusion}

We have emphasized the existence of two astrophysical fractals,
the interstellar medium, with structures ranging from 10 AU to
100 pc, and the large-scale structures of galaxies, from 10 kpc
to 200 Mpc at least. The first one is in statistical equilibrium,
while the second one is still growing to larger scales. In both
cases, we can describe these media as developping large-scale
fluctuations with large correlations as is familiar in critical
phenomena. We have investigated the hypothesis 
that in both cases, self-gravity is the main force
governing these fractal structures.

Numerical simulations can help to understand the formation
of these structures, and the main mechanisms at play.
Unfortunately, practical constraints confine the possibilities
to a limited range of scales, and results are often ambiguous.
Simulations of MHD turbulence with self-gravity are not yet
to the point to reproduce the scaling relations observed 
in the ISM. Purely self-gravitating simulations without
large-scale injection of energy produce only transient
fractal structures, depending in a large part on initial
conditions. When large-scale energy is taken into
account (by the galaxy shear namely), a quasi-stationary
fractal structure, over $\sim$ two orders of magnitudes,
can be obtained. This is encouraging, waiting for
more performant 3D simulations.

A statistical thermodynamic approach of
self-gravitating systems has been developped, and
it is shown that the phenomenological potential, which is
in $1/r$ between two cutoffs (at small and large-scale),
can be described by a scalar field theory.
 Using renormalization group methods, the system
is found to be of the  same
universality class as the Ising $d=3$ model.
The critical exponents can then be derived,
and the fractal dimension $D=1.6$ deduced.

The gravitational gas appears to be critical for a large range 
of temperatures and couplings, while for spin models there is
only a critical value of the temperature. This feature must be 
connected with the scale invariant character of the Newtonian force
and its infinite range.

\end{document}